\documentclass{elsart}

\usepackage{epsfig}
\usepackage{amssymb}
\usepackage[applemac]{inputenc}
\usepackage{amsbsy}

\begin{document}

\begin{frontmatter}

\title{\bf{Growth Optimal Investment and Pricing of Derivatives}}
\author[EA]{Erik Aurell\thanksref{thanksEA}},
\author[RB]{Roberto Baviera},
\author[EA]{Ola Hammarlid},
\author[RB]{Maurizio Serva} and
\author[AV]{Angelo Vulpiani\thanksref{thanksAV}}

\address[EA]{Department of Mathematics, 
Stockholm University, S-106 91 Stockholm, Sweden}
\address[RB]{I.N.F.M., Unit\`a dell'Aquila, Via Vetoio 1
I-67010 Coppito, Italy}
\address[AV]{I.N.F.M., Unit\`a di Roma ``La Sapienza'',
P.le Aldo Moro 2 I-00185 Roma, Italy}
\thanks[thanksEA]{This work was supported by the NFR contract S-FO 1778-302}
\thanks[thanksAV]{and by I.N.F.M. grant-in-aid}.

\begin{abstract}
We introduce a criterion how to price derivatives
in incomplete markets, based on
the theory of growth optimal strategy in repeated multiplicative games.
We present reasons why these growth-optimal strategies
should be particularly relevant to the problem of pricing derivatives.
Under the assumptions of no trading costs,
and no restrictions on lending, 
we find an appropriate 
equivalent martingale measure that prices 
the underlying and the derivative
security. 
We compare our result with other alternative pricing procedures
in the literature, and discuss the limits of validity of
the lognormal approximation. We also generalize the pricing method to a market 
with correlated stocks. The expected estimation error of the 
optimal investment fraction is derived in a closed form,              and its validity
is check  with a small-scale empirical test.  
\end{abstract}
\end{frontmatter}

\newpage
\section{Introduction}
\label{s:introduction}

The option pricing formula of Black-Scholes is arguably one of
the most successful modern applications of quantitative mathematical analysis. 
Despite its great simplicity
it gives prices of the most
actively traded options on
leading markets which are typically not off by more than a few percent.
This has paved the way for an influx
of advanced ideas from the theory of stochastic processes
into finance.
In the most general formulation of the no-arbitrage argument
of Black-Scholes~\cite{BlackScholes} and Cox et al~\cite{CoxRossRubinstein}
it was established
by Harrison-Kreps~\cite{HarrisonKreps} that
a price system (under certain
restrictions such as no trading costs) admits no arbitrage opportunities
if and only if there exists an equivalent probability measure, 
with respect
to which all discounted 
price processes of underlying and derivative securities
are martingales.

This theory is presented in several
excellent monographs, see~\cite{CoxRubinstein,Ingersoll,Merton,Duffie,Hull}.
Nevertheless, these developments do not yield a fully
general solution to the derivative pricing problem, since the equivalent
martingale measure is not uniquely defined, except if the market
is complete.
If the market is incomplete there are many possible equivalent
martingale measures, and each of these specifies a price system
without arbitrage opportunities.

In the lognormal model
of Black-Scholes the market of one share and one bond is
already complete. One can thus test its predictions by
computing the volatility of
the price process of the underlying security, plugging
that into the Black-Scholes formula, and then comparing with observed
market prices of a derivative: such tests were initiated already 
by Black-Scholes~\cite{BlackScholes72}
and Black~\cite{Black75}.
The volatility from historical data is obviously
subject to measurement errors
and statistical fluctuations. A more robust test is therefore
if observed option prices can be fitted by a Black-Scholes formula
with volatility treated as a free parameter. There are several recent reports
that this is not so, see Rubinstein~\cite{Rubinstein1994},
and earlier investigations~\cite{MacbethMerville,Rubinstein1985,Chance}.

In addition to the lognormal model there exists
a wider class 
of It\^o processes in continuous time
for which the market is also complete, and the derivative pricing
problem can be solved only by no-arbitrage conditions.
To distinguish which model in this class is the best approximation to observed
data, and then to use it, is however
already not a trivial task, see 
e.g~\cite{DermanKani,DermanKani1998,Rubinstein1994,Bates1995,Bates}.

For the rest of this paper we will 
consider the general case of 
an incomplete market, but we do keep the standard assumption
of no market friction.
We introduce a criterion to price derivatives
in this situation.
Stated differently, we advance an argument as to which out of
many possible equivalent martingale 
measures is the appropriate one. 
We refer to this criterion as the Principle of No Almost Sure
Arbitrage~\cite{AurellBavieraHammarlidServaVulpiani}: it is based on the theory of optimal gambling 
of Kelly~\cite{Kelly}. We will consider an investor that
gambles on investing in an underlying and a derivative security,
and who looks for the growth-optimal
investment strategies among such combinations.
 
The paper is organized as follows:
in section~\ref{s:kelly} we summarize 
and discuss the Kelly theory in a version
appropriate for our purposes. 
We discuss also here the papers
of Samuelson~\cite{Samuelson} and Merton-Samuelson~\cite{MertonSamuelson}.
In section~\ref{s:principle-NASA},
the technical core of the paper, we apply the Kelly theory to
derivative pricing, and in section~\ref{s:remarks} we discuss
the relations and differences to other proposed
alternative derivative pricing procedures.  

It is straight-forward to see that in complete
markets our procedure agrees with the standard ones, e.g. with the models
of Cox-Ross-Rubinstein and Black-Scholes.
In section~\ref{s:lognormal} we discuss the lognormal limit of our
procedure, the limits of its applicability, and that it is not
exactly equivalent to the Black-Scholes formula.
In section~\ref{s:error} we study the expected scatter of the 
 optimal investment fraction and report a small-scale empirical test  
 of pricing options on the Swedish OMX index.
 In section~\ref{s:multiasset} we generalize the 
method to a market with correlated stocks.  
In section~\ref{s:conclusion} we summarize and discuss 
our results.

An earlier version of this paper was circulated in 1998~\cite{Aurelletalshort}.
A longer sequel was also 
made available in the electronic 
archive~\cite{AurellBavieraHammarlidServaVulpiani}, 
lacking however in the version accepted for publication, 
material corresponding to sections~\ref{s:error}
and ~\ref{s:multiasset} below. The presentation in the present paper 
is significally different from 
~\cite{AurellBavieraHammarlidServaVulpiani} and contains  a fuller 
discussion of the implementation of the model and the economic 
background. 

\section{The Kelly problem of Optimal Gambling}
\label{s:kelly}
The theory to be exposed in this section is due to Kelly~\cite{Kelly},
who was looking for an interpretation of Information Theory of Shannon \cite{Shannon}
outside of the context of communication, and
to the treatment of the
St.~Petersburg Paradox by Bernoulli~\cite{Bernoulli}.
For a recent review of growth-optimal investment strategies,
see Hakanson-Ziemba~\cite{HakansonZiemba}.

Consider a price movement of stock or some other
security which is described by
\begin{equation} 
S_{i+1}=u_i S_{i}
\label{eq:u_i-definition}
\end{equation} 
where time is discrete, $S_i$ is the price at time $i$ and
the $u_i$'s are independent, identically distributed
random variables. 
The assumptions on the $u_i$'s can be relaxed.

Consider now an investor which starts at time $0$ with a
wealth $W_0$, and who decides to gamble on this stock repeatedly.
Suppose that the investor chooses to commit at each time
a fraction $l_s$ of his capital in stock,
and the rest in a risk-less security.
We set the risk-less rate to zero.
A non-zero risk-less rate
corresponds to a discount factor
in the definition of the share prices, and can be accounted for by
a redefinition of the $u_i$'s.
The investor will hence
hold at time $i$ a number $l_s W_i/S_i$ of shares, and his wealth
at successive instants of time follows
a multiplicative random process
\begin{equation} 
W_{i+1}=(1-l_s)W_i + l_su_i W_i = \left(1+l_s(u_i-1)\right)W_i
\label{eq:W_i-definition}
\end{equation}
In the large time limit we have by the law of large numbers
that the exponential growth rate of the wealth tends 
with probability one to a constant.
That is,
\begin{equation}
\lambda(l_s) = \lim_{T\to\infty}\frac{1}{T} \log\frac{W_T}{W_0} 
= E^p[\log\left(1+l_s(u-1)\right)].
\label{eq:lambda-definition}
\end{equation}

The optimal gambling strategy in the sense of Kelly consists in maximizing
$\lambda(l_s)$ in (\ref{eq:lambda-definition}) by varying $l_s$.
The solution must be unique because the logarithm is a concave function of
its argument.
Which values of $l_s$ are reasonable in our problem?
First, the optimum $l_s$ must be such that $1+l_s(u-1)$ 
is positive on the support of $u$.
Second, we must decide if the investor is allowed to borrow money.
In the original formulation of Kelly he is not, but here it is useful
to assume that the investor may do so at a risk-less rate.
Furthermore we also assume that the investor can go short i.e., it is 
possible to borrow stocks.  Hence we allow $l_s$ to take any finite
positive or negative value, and look for the maximum of 
$\lambda(l_s)$.

The desired strategy is hence the only finite $l_s$ which solves
\begin{equation}
 \frac{d \lambda(l_s)}{dl_s}|_{l_s={l_s}^*} 
= E^p[\frac{u-1}{1+{l_s}^*(u-1)}]=0
\label{eq:l-star-definition}
\end{equation}
and the maximum realized growth rate is
\begin{equation}
\lambda^* = E^p[\log\left(1+{l_s}^*(u-1)\right)]
\label{eq:lambda-star-definition}
\end{equation}
Let us look more closely at
(\ref{eq:l-star-definition}) if $u$ can take the values
$v_n$ with probabilities $p_n$.
Let us write 
\begin{equation}
q_n = \frac{p_n}{1+l_s^*(v_n-1)}
\label{eq:q-definition}
\end{equation}
such that (\ref{eq:l-star-definition}) reads
\begin{equation}
\sum_n q_n (v_n-1) = 0
\label{eq:q-equation}
\end{equation}

From $1+{l_s}^*(v_n-1)$ being positive it follows that the $q_n$'s
are all positive, and from (\ref{eq:q-equation}) it follows that
the sum of the $q_n$'s is one. The $q_n$'s thus form a new set
of probabilities, and, again by (\ref{eq:q-equation}),
with respect to these the stock
price process is a martingale.

It is worthwhile to discuss this point at some length in light of the
well-known papers of Samuelson~\cite{Samuelson} 
and Merton-Samuelson~\cite{MertonSamuelson}. 
These authors demonstrated
that if the risk/return preferences of an agent is described
by an expected utility, then the growth-optimal strategy does not
maximize that utility, except if the utility 
is logarithmic, and the price
process is similar to~(\ref{eq:u_i-definition}).

It is a mathematical fact that the growth
rate~(\ref{eq:lambda-definition}) holds with probability one.
If a gambler would choose any other strategy 
in $l_s$ but that which maximizes
$\lambda$ then he would in the large time limit
almost surely end up with an exponentially smaller capital.

In the counter-examples of Samuelson-Merton
the dominant
contribution to non-logarithmic utility comes from
events with asymptotically (in time) vanishing probability.
When 
sets of measure zero are involved, the relation between
sample and ensemble averages becomes delicate.
Suppose some agents want to maximize a 
non-logarithmic utility, and that they choose strategies
accordingly. 
If they do so, and we compare with them using the growth-optimal
strategy, they would almost surely
end up with less utility according to their own criterion.
Only if there would be an ensemble of infinitely many 
agents, all using this same strategy, then some few would
actually end up with higher utility. And, in fact,
so much higher
that the averaged utility over the ensemble would be increased.
This is the sense in which other strategies 
can be `better' than the growth-optimal strategies.
Therefore, Merton-Samuelson give preference to ensemble averages 
over sample average. This is a problematic point, which has apparently 
not received much attention in the economic literature. 

A simple counter-argument against the growth-optimal criterion is 
that time in economical problems, in the sense considered here, is
perhaps often not very long. However, this is essentially a quantitive 
question, which has to be decided case-by-case. For recent discussions
of characteristic times in the Kelly problem with transaction costs, 
see  ~\cite{AurellMuratore,Baviera,Serva}

\section{Fixing the price of the derivative}
\label{s:principle-NASA}
Let us now consider the stock of section~\ref{s:kelly} and
one derivative security on the same stock.
The argument generalizes at every step in an obvious
way to an arbitrary number of derivatives, but to 
keep the notation simple
we will write out the formulae for
one derivative.
To begin, let us write the present share price at the  time $i$ 
$S_{i}$, and let us assume that 
the realized value of the derivate
at the next future instant of time
is $C_{i+1}(S_{i+1})=C_{i+1}(S_{i}u_{i})$.
The unknown parameter is the present price of derivative, which we
will write $C_i$.
We will further suppose that this situation is repeated over and over again.
That is, we assume
that the investor finds himself many times in the situation
that if he keeps some money $W_{i}$ in stock it will be worth
$u_{i}W_{i}$, while if he keeps the same money in the derivative it will
be worth $C_{i+1}(S_{i}u_{i})W_{i}/C_{i}(S_{i})$, 
with the same random variable $u_{i}$ entering
into both expressions. The simplest example 
is one time step before 
expiration time of a call option
with a present share price $S_{T-1}$ and a strike price
$K$; in this case one has 
$C_{T}(S_{T})=max(S_{T}-K,0)$.

If at the $i$'th stage in this gamble the
investor commits a fraction $l_s$ of his wealth in shares and
a fraction $l_d$ in the derivative, this means $l_sW_i/S_i$ shares and
$l_d W_i/C_{i}$ derivatives.
After the random variable $u_i$ has taken a value,
the wealth is changed to
\begin{equation}
W_{i+1} = \left(1 + l_s(u_i-1) +
l_d(\frac{C_{i+1}}{C_{i}}-1)\right)W_{i}
\label{eq:double-portfolio}
\end{equation}
If the investor plays this game many times, his wealth will almost
surely grow at an exponential rate (over these games) 
which is
\begin{equation}
\lambda(l_s,l_d;C_i)
= E^p[\log\left(1+l_s(u-1)+l_d(\frac{C_{i+1}}{C_{i}}-1)\right)]
\label{eq:double-Lyapunov-exponent}
\end{equation}

Let us now assume that we compute (\ref{eq:double-Lyapunov-exponent})
and that we find an optimal growth rate
\begin{equation}
\lambda^*(C_i)
= \hbox{Max}_{l_s,l_d}
 E^p[\log\left(1+l_s(u-1)+l_d(\frac{C_{i+1}}{C_{i}} - 1)\right)]
\label{eq:double-Lyapunov-exponent-optimal}
\end{equation}
and that this is realized at fractions $l_s^*$ and $l_d^*$.
If $l_d^*$ is larger than zero, then  all operators
would like to buy the derivative. Hence its present price $C_{i}$
would tend to rise, and its rate of return would fall. The fraction
$l_d^*$ would thus tend to fall. If, on the other hand, $l_d^*$  would
be less than zero, then all operators would want to go short on the
derivative, its present price would fall, its rate of return would
rise and $l_d^*$ would tend to rise.
The only value of $C_i$ in which the market can be in equilibrium is 
therefore the one such that $l_{d}^{*}$ is zero. This statement was 
referred to as the Principle of No Almost Sure Arbitrage in 
\cite{AurellBavieraHammarlidServaVulpiani}.
We remark that 
from the technical point of view,
this procedure, where $l_d^*$ is taken to vanish, is closely similar
to the Samuelson's~\cite{SamuelsonWARRANTS} methodology of warrant pricing 
in the `incipient case', see Merton~\cite[chapter~7]{Merton}.

 $C_{i}$ and the fractions $l_s$ 
and $l_d$ thus satisfy two equations. One is
\begin{equation}
0\, =\, \frac{d \lambda(l_s,l_d;C_{i}) }
{dl_s}|_{l_s=l_s^*,l_d=0,C_{i}=C_{i}^{*}}\, =\,
 E^p[\frac{u-1}{1+l_s^*(u-1)}]
\label{eq:l_s-star-definition}
\end{equation}
which is identical to 
(\ref{eq:l-star-definition}) of the Kelly model, and therefore
determines $l_s^*$.
The other is
\begin{equation}
0\, =\, \frac{d \lambda(l_s,l_d;C_{i}) }
{dl_d}|_{l_s=l_s^*,l_d=0,C_{i}=C_{i}^{*}}\, =\,
 E^p[\frac{(\frac{C_{i+1}}{C_{i}^{*}}-1)}{1+l_s^*(u-1)}]
\label{eq:C_0-star-definition}
\end{equation}
If we now compare with 
(\ref{eq:q-definition}) and (\ref{eq:q-equation})
we see that (\ref{eq:C_0-star-definition}) can be written
\begin{equation}
C_{i}^{*} = \sum_n q_n C_{i+1}(S_{i}v_n)
\label{eq:equivalent-Martingale-pricing}
\end{equation}
which states that the derivative price process is a martingale
with respect to $q$.

We have constructed the equivalent martingale measure $q$ over
one elementary time-step. 
By compounding we can construct, 
from $T$ convolutions of $q$ distribution, the price
distributions over many time-steps $Q_T(S_T|S_0)$. 
The price of, say, a European call option with expiration time $T$ at time
$t$, can thus be written in the standard manner as
\begin{equation}
C(i,S_i) =  E^{Q_{T-i}}[C(T,S_T)]
\quad C(T,S_T) = max( S_T-K,0)
\label{eq:equivalent-Martingale-pricing-T}
\end{equation}
where the expectation value is taken with respect 
to the many time-steps distribution $Q_{T-i}(S_T|S_i)$
over share prices $S_T$
at time $T$, conditioned by the share price having been
$S_i$ at time $i$. We have taken the risk-less rate to be zero.
A non-zero risk-less rate can be put back in as a discount factor
in the share prices, which leads to a different definition of the
$u$'s, the returns over elementary periods of time, and hence to
different $q$'s. 

\section{Comparison with other derivative pricing prescriptions}
\label{s:remarks}
The pricing procedure proposed in section~\ref{s:principle-NASA}
is based on a particular choice of an equivalent martingale
measure. It therefore follows that this procedure is arbitrage-free.
It also follows that it must agree with no-arbitrage pricing 
in complete markets.

It may nevertheless be instructive to carry through the calculations
of section~\ref{s:principle-NASA} explicitly for the
Cox-Ross-Rubinstein model. We hence assume that the 
discounted share price
can go up by a fraction $u/r$ with probability $p$, 
down by $d/r$ with probability
$(1-p)$, and that $u>r>d$,
where $r$ is the risk-free interest rate.
Equation~\ref{eq:l_s-star-definition} 
is then solved for $l_s^*$ as
\begin{equation}
l_s^* = \frac{rp}{r-d} - \frac{r(1-p)}{u-r}
\quad ; \quad
q=\frac{p}{1+l_s^*(\frac{u}{r}-1)}
= \frac{r-d}{u-d}
\end{equation}
and this reproduces the Cox-Ross-Rubinstein measure $q$, independent of $p$.
The maximum growth rate however still depends on $p$, and is
\begin{equation}
\lambda^* = -p\log\frac{p}{q} - (1-p)\log\frac{1-p}{1-q}
\label{eq:CRR-Kullback-contrast} 
\end{equation}
The Black-Scholes model can now be derived as the continuous-time limit
of the Cox-Ross-Rubinstein model.

The growth rate (\ref{eq:CRR-Kullback-contrast}) is the Kullback contrast
of the measure $q$ with respect to the measure $p$
in the dichotomic case. 
It is easy to
see that this is the general form of the solution,
and that the measure $q$ can be 
obtained as follows: Consider a security with price process determined
by a measure $p$, and consider all measures $q$ with respect to which the
price process of the security is a martingale.
Then the particular measure $q$ which we compute from
(\ref{eq:l_s-star-definition}) (see also
equations~(\ref{eq:q-definition}) and~(\ref{eq:q-equation}))
is the one that minimizes the Kullback contrast
\begin{equation}
d_p(q) = E^p [-\log\frac{p}{q}] = 
-\sum_n  p_n \log\frac{p_n}{q_n} 
\label{eq:Kullback-contrast}
\end{equation}
Our proposal therefore coincides, over one time step, with one of the
two proposals recently put forward by Stutzer~\cite{Stutzer}.
Stutzer's formulae differ from ours in that we posit
the minimization of
(\ref{eq:Kullback-contrast})
under the martingale constraint
for the elementary process over each
time-step, while Stutzer applies (\ref{eq:Kullback-contrast})
directly on the probability distribution over a finite time
interval. A similar approach minimizing the relative entropy distance 
to calibrate a pricing model has been suggested by Avellaneda et al 
\cite{Avellaneda}.

It is also interesting to remark that if we perform 
the quadratic 
approximation of the logarithm in (\ref{eq:lambda-definition}) 
around 
the expected return on capital, $E^p[W_{i+1}/W_i]$,  
and then look for the optimal portfolio, we obtain 
\begin{equation}
q(u) = \left( 1 - \frac{\mu}{\hbox{Var}^p[u]}
(u-1-\mu)\right) p(u) \quad ; \quad \mu = E^p[u-1]
\label{eq:quadratic-expansion}
\end{equation}
Equation (\ref{eq:quadratic-expansion}) reproduces the least-square option
pricing procedure of Follmer-Sondermann~\cite{FollmerSondermann},
Sch\"{a}l\cite{Schal} and Schweizer~\cite{Schweizer}. 
Another approach to option pricing based on global least-square
minimization was recently proposed
by Bouchaud-Sornette~\cite{BouchaudSornette}, 
see also~\cite{BouchaudIoriSornette}, but has
now been proven by Wolczynska \cite{Wolczynska_Article}
and Hammarlid~\cite{Hammarlid} to give the same least-square solution, see
also Schweizer~\cite{Schweizer2}.
The empirical tests of~\cite{BouchaudSornette} and~\cite{BouchaudIoriSornette},
which pertain mainly to the
case of small $\mu$, hence
also validate our theory.

A conceptual problem of least-square option pricing,
first shown by Dybvig-Ingersoll~\cite{DybvigIngersoll},
is that it can assign negative prices of some state-contingent
claims. This may lead to negative prices of derivatives
with strictly
non-negative pay-off, and hence to arbitrage opportunities.
In (\ref{eq:quadratic-expansion}) this happens when $u$ is sufficiently
large, provided $\mu$ is positive.
From our point of view this result is clear. 
Negative prices appear only at capital values for which 
the quadratic approximation is a decreasing
function. It
is precisely because quadratic utility is not monotonically
increasing that its maximization only gives a pseudo-martingale
measure, which can take both positive and negative values.
Our theory is practically equivalent to
the least-square method when that one has no problems, i.e., when
we look at small deviations from the most probable realizations
of the capital returns and when $\mu$ is small.
In the opposite limits our theory deviates from 
least-square minimization, always prices derivatives by a proper
equivalent martingale measure, and the problem of possible negative prices
does not appear.

\section{On the lognormal limit}
\label{s:lognormal}

The distribution $Q_T(S_T|S_0)$ is constructed by compounding the measure
 $q$ of one time-step.
If $u$ only takes only a finite  number  of values,
then $Q_T(S_T|S_0)$ can be written as a multinomial
expansion.
However, if the number of time steps is large,
then the multinomial expansion is unwieldy, and some other
approach must be found for analytic and numerical work.

The asymptotic price distribution is of course determined by the behavior 
of aggregated logarithmic returns, and this is intimately linked to
the lognormal distribution, its applicability and its limits. 
Let us assume that  $E^q [ \left( \log u \right)^2 ]$ is finite.
Then, for the most probable realizations,
$Q_T(S_T|S_0)$ is well approximated by 
a lognormal distribution, i.e. 
\begin{equation}
Q_T(S_T|S_0) \simeq
 Q_T^{(LN)}(S_T|S_0) =
 {1 \over {S_T \sqrt{2 \pi  \Delta^2 T} }}
  \exp - { {(\log (S_T/S_0) - \lambda T)^2}  \over {2  \Delta^2 T} } 
\label{eq:lognormal-approximation}
\end{equation}

where 

\begin{equation}
\lambda=E^q[\log u] \quad ;  \quad
\Delta^2= E^q[(\log u)^2]-(E^q[\log u])^2
\label{eq:lambda-delta-definitions}
\end{equation}
For large $T$ only low-probability events are not distributed
according to~(\ref{eq:lognormal-approximation}).
For the treatment of  such large, but very rare, events
the proper mathematical setting is the
theory of large deviations, see Varadhan~\cite{Varadhan}.
In ~\cite{AurellBavieraHammarlidServaVulpiani}
we discuss the use of large deviations theory to price
derivatives related to rare events.

We note  that $ Q_T(S_T|S_0)$ 
is by construction a martingale measure, while, 
for the log-normal approximation,
\begin{equation}
 E^{Q^{(LN)}}[S_T]=\int Q_T^{(LN)}(S_T|S_0) \, S_T \, d S_T = 
  S_0 \exp \, (\lambda+{1 \over 2} \Delta^2) T \quad .
\label{eq:lognormal-martingale-condition}
\end{equation}
In general  $ \lambda+{1 \over 2} \Delta^2 \ne 0$. That is, 
in the lognormal approximation
one looses the martingale property.

Our lognormal approximation
does not agree with a straight-forward
application of the Black-Scholes formula.
We recall that the Black-Scholes theory also involves 
lognormal distribution.
In the parameterization of (\ref{eq:lognormal-approximation})
it is characterized by drift coefficient $\lambda_{BS}$
and volatility $\Delta_{BS}^2$, related through
\begin{equation}
\lambda_{BS}=-{ 1 \over 2} \Delta_{BS}^2 \, , \,\,\,
 \Delta_{BS}^2=E^p[(\log u)^2]-(E^p[\log u])^2
\end{equation}
It is easy to understand that if $u$ has support concentrated around
one, then the difference between $Q_T^{(LN)}$ and $Q_T^{(BS)}$
is small.
In the continuous limit the differences between $Q_T$, $Q_T^{(LN)}$ 
and $Q_T^{(BS)}$ disappear altogether.

\begin{figure}
\begin{center}
\mbox{\psfig{file=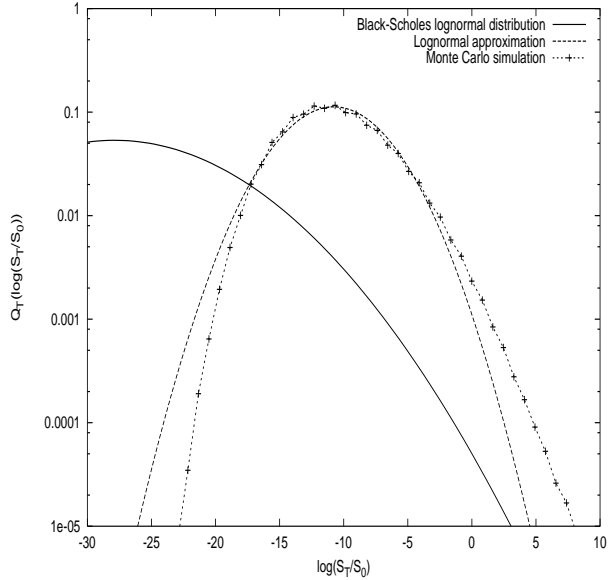,height=8cm,width=8cm,angle=270}}
\end{center}
\caption{
The probability distribution is a discrete approximation
to a L\'evy distribution.
The relative price increments can take the values
$u_n=u_0 \alpha^n$, with $u_0$ equal to $0.4$,
$\alpha$ equal to $1.2$
and the index $n$ ranging from zero to $N$,
which in this example has been set to $30$.
The probability of an elementary event,
$p(u_n)$, is $C_N u_n^{-\beta}$, with $\beta$ equal
to $0.5$ and $C_N$ a normalization constant.
The Monte Carlo simulation was performed with $20$
million trials.
}
\label{f:compared-distributions-farwest}
\end{figure}

On the other hand, the risk-neutral price distribution can 
actually be close to lognormal but still 
significantly different from
Black-Scholes.
In an extreme situation of very large $Var^q[u]$,
$Q_T^{(LN)}$ is only close to $ Q_T$ in a very limited region.
As an example we show in figure~\ref{f:compared-distributions-farwest}
the case where
$p(u)$ is a truncated 
Levy distribution. The distribution
$Q_T$ has been
computed by a Monte Carlo procedure.
Here it is obvious that one has to consider the corrections to the lognormal
to get a good approximation to $Q_T$.

\begin{figure}
\begin{center}
\mbox{\psfig{file=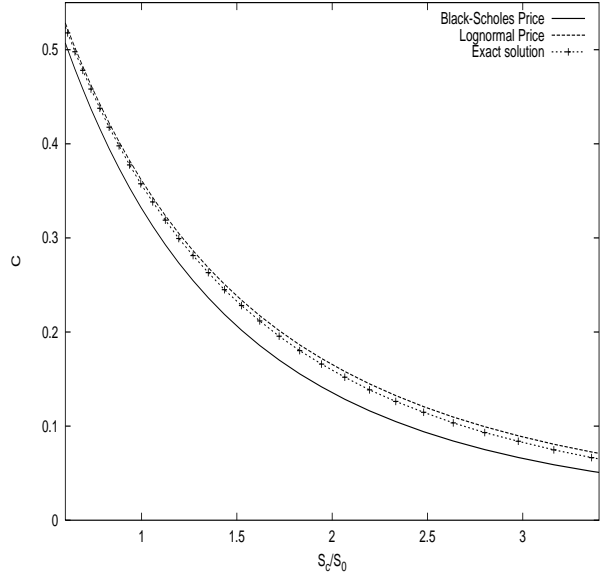,height=8cm,width=8cm,angle=270}}
\end{center}
\caption{
Option prices $C$ as function of the strike prices $K$ for $T=30$.
The relative price increments over one
elementary step in the process are, that 
$u$ takes three discrete values $0.8$, $1$ and  $1.2$ 
with probabilities $0.2$, $0.3$ and $0.5$.}
\label{f:prices-narrow}
\end{figure}

We observe that if $u$ has support around one
the predictions of the Black-Scholes theory
and the lognormal approximation
are both rather close
to the exact formula: this is the case
in figure~\ref{f:prices-narrow}.
In figure~\ref{f:prices-narrow} we show
the price of a European call
option as a function of strike price
where the probability distributions
are: the increment $u$ takes three discrete values $0.8$, $1$ and $1.2$ 
with probabilities $0.2$, $0.3$ and $0.5$.

On the other hand, in general,
both Black-Scholes formula and the lognormal approximation 
can be far from the exact formula. 
Large deviations have therefore to be taken into account
to compute properly the expectation value,
see~\cite{AurellBavieraHammarlidServaVulpiani}.

\section{The experimental error on the optimal $l^*$}
\label{s:error}
\noindent
The realistic application of the theory we have presented
depends on the possibility of estimating
the distribution $p(u)$ from real data.
In fact, $p(u)$ enters into the construction
of the $q(u)$ both directly and with the optimal $l^*$,
which is a functional of $p(u)$.
Therefore, it is very important to estimate $l^*$ together
with the magnitude of its error.

The information which is possible to obtain from the market is a 
series of stock prices, $S_{0}, \ldots S_{H}$,
where $H$ is the number of data. 
We consider the time interval between observations 
to be large enough to ensure the time independence
of the returns $u_i = S_{i}/S_{i-1}$.
 
In order to estimate experimentally the true distribution $p(u)$ 
we consider a partition of the support of the distribution in 
$M$ intervals of width $\Delta $, the requirement being
that  $M \leq H$. 
Then we estimate the probability in one of the interval
by the ratio between the number of observed data in the interval 
and the total number of data $H$.
In this way one creates a histogram of the approximating
probability  $P_{\Delta,H}( u)$, the  deviation of which from
the `true' probability $p(u)$ decreases as $H$ becomes larger and
$\Delta$ smaller.  
Let us call this deviation
\begin{equation}
	\delta P_{\Delta,H} (u)=P_{\Delta,H} (u)-P(u) \,\, .
	\label{dP}
\end{equation}
Because an observed data is either in an interval or not, 
the number of observed data in an interval follows a binomial
distribution.
This observation immediately leads to
\begin{eqnarray}
	E[\delta P_{\Delta,H}(u)]&=&0
	\label{e}  \\
	Var(\delta P_{\Delta,H}(u)) & =&  P_{\Delta,H}(u)
     (1-P_{\Delta,H}(u)\Delta)
	\label{var}  
\end{eqnarray}
To estimate the true $l^{*}$ with  $l^*_H$ using the
approximating probability $P_{\Delta,H}(u)$ 
we have to solve the equation
\begin{equation}
\int \frac{P_{\Delta,H}(u) (u-1)}{1+l^*_H(u-1)} du=0\,\, .
	\label{estdlambda}
\end{equation}

The error on the optimal fraction invested in the stock
is the difference:
\begin{equation}
	\delta l=l^*_H-l^{*},
	\label{dl}
\end{equation}
where $l^*$ is the true optimal fraction obtained from
the distribution $p(u)$.

A Taylor expansion up to the second
order around $l^*$ of (\ref{estdlambda}),
and the above considerations about 
the error in estimating the probability give 
after some calculations
\begin{equation}
  \sigma_l = \frac{1}{\tilde{\sigma}\sqrt{H}}\,\, .
	\label{SDl}
\end{equation}
where
\begin{equation}
\tilde{\sigma} \equiv \left( \int 
\frac{ P_{\Delta,H}( u)(u-1)^2}{(1+l^*(u-1))^2}du\right)^{\frac{1}{2}}
\end{equation}
where it has been used the fact that for a sensible
partitioning of the sample space $P_{\Delta,H}(u) \Delta \ll 1$.

\begin{figure}
\vspace*{13pt}
\begin{center}
\mbox{\psfig{file=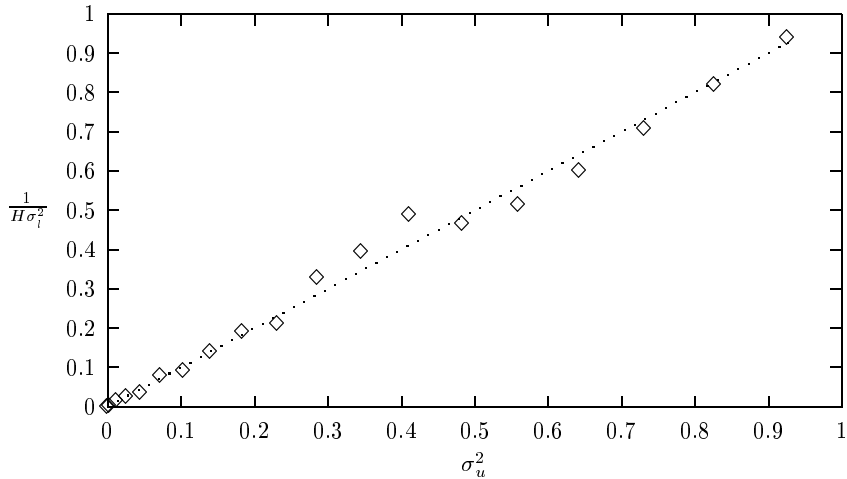,width=4.5in}}
\end{center}
\vspace*{13pt}
\caption{
Comparison between the error on $l^*$ estimated by 
equation~(\ref{SDl}) and by the bootstrap technique.
We plot $1/(H \sigma_l^2)$ {\it vs} $\sigma_u^2$, 
where the $\sigma_l$ are obtained by the bootstrap technique with 
$J=200$ samples and
the distribution of $u$ is a gaussian with $E[u]=1$ and 
standard deviation 
$\sigma_u$.
}
\label{fig:Ola}
\end{figure}

The error on $l^*$ can also be estimated
from the historical distribution using bootstrap technique~\cite{Efron},
i.e. drawing with replacement a set of $J$ new samples of the same size $H$
of the original one from the empirical distribution,
and computing a new $l^*$ for each bootstrap sample.
In figure~\ref{fig:Ola} we compare the variance of this distribution
with the quadratic estimation~(\ref{SDl}).
The agreement is very good for different values of the 
$\sigma_u \equiv \sqrt{Var[u]}$.

According to equation~(\ref{SDl}), 
it is evident that in most cases the error of $l^*$ is large. 
A non-intuitive result is that it is easier to determine $l^*$ with accuracy
for distribution with larger variance 
because of the $\tilde{\sigma}$ in the denominator.
This fact can be understood as follows. A stock with small
fluctuations allows for a bigger fraction be held against the interest rate
to make money, but just a small change in the distribution will make a
big change in the optimal fraction. 
On the contrary, for a wildly fluctuating stock the fraction
invested is smaller, and a little change in the distribution will not
have much effect on the optimal portfolio.

\subsection{Data analysis}
\noindent
Empirical study is needed to verify the growth optimal pricing
procedure here discussed.
To verify the model we looked at the
European call option on  the OMX-index, traded at the OM exchange in 
Stockholm, during the period 2 January 1995 until 22 September 1995.
We consider closing prices of each day. The time period between
trading is one day. The time series contained
$183$ observations, providing a sample of $u_{i}$ of size (data points)
$182$.
The empirical distribution of $u_i$ of the OMX-index
is built by the fraction of the price of
the OMX-index today and tomorrow, $u_{i}=S_{i+1}/S_{i}$.
Weekends and other holidays are treated as non-existing.

  The thirty day interest
rate was as used discount factor $r$ to compute the 
discounted daily returns, $\{u_{i}/r_{i}\}^{182}_{i=1}$. 
During the period the interest rate fluctuate between,
$7.15-9.83 \%$ of yearly yield. 
The sample space was divided into $M=H$ cells and
the optimal $l^{*}$ determined 
using the Newton-Raphson algorithm~\cite{NumericalRecipes}. 
Because of the concave of the logarithm the convergence
is very fast, the optimal $l^*$
for the OMX-index is equal to $14.46$.

The distribution of $l^*$ is obtained using bootstrap technique
with a set of $200$ new samples.
The standard deviation of $l^{*}$ is $8.6$
and can be compared to $8.7$
obtained
estimating the standard deviation using formula~(\ref{SDl}).
From the empirical distribution $p$ the empirical $q$-distribution
can be created and from  $q$-distribution the empirical $Q$-distribution 
can be derived.  

The empirical $q$-distribution is used to estimate parameters of the
lognormal approximation~(\ref{eq:lambda-delta-definitions})
$\lambda=1.9\cdot 10^{-4}$ and $\Delta^2=7.3\cdot 10^{-5}$.
We determine, using the bootstrap technique, the distribution and
standard deviation for
$\lambda$ and $\Delta^2$, 
 $\sigma_{\lambda}=3.6 \cdot 10^{-6}$ and
 $\sigma_{\Delta^2}=3.9\cdot 10^{-6}$ respectively.

These parameters correspond, inside the experimental errors,
to the Black-Scholes model.
The consequence is
that in this case lognormal approximation of the
Kelly pricing scheme is practically equivalent 
to Black-Scholes model. 

\begin{figure}
\vspace*{13pt}
\begin{center}
\mbox{\psfig{file=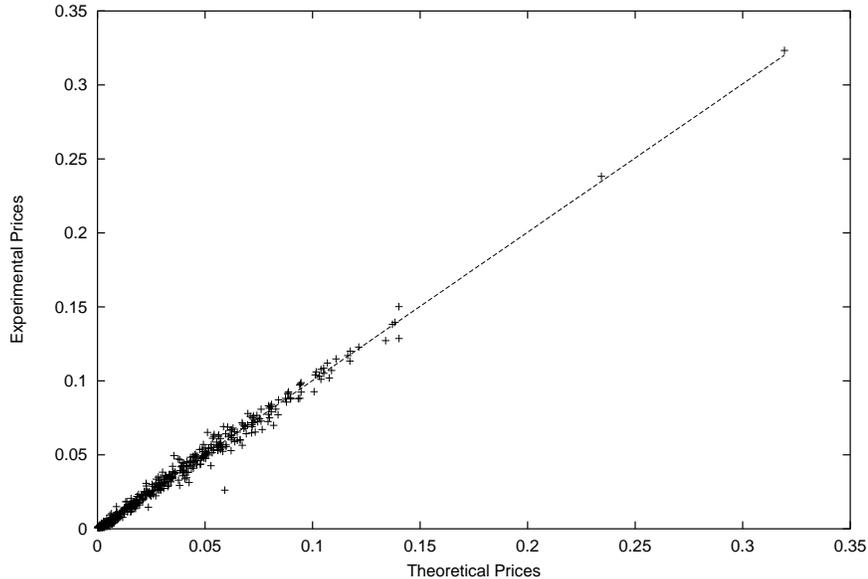,angle=270,width=4.5in}}
\end{center}
\vspace*{13pt}
\caption{
Experimental option prices against 
theoretical ones for the call options of OMX with maturity date less
or equal then $10$ working days in the period between 
the $2^{nd}$ of January and the $22^{nd}$ of September 1995. 
The theoretical prices are generated with the 
probability $Q_T(S_T/S_0)$ using the  
Monte Carlo discussed in this section with $10^5$ trials.
The linear fit with $a x +b$ gives $a=1.003 \pm 0.004$
and $b=0. \pm 10^{-4}$.
As discussed in the text the major source of errors is 
on the estimation of $l^*$ due to the low data frequency considered.
}
\label{fig:Th_Exp}
\end{figure}

The comparison between empirical data 
and theoretical one is shown in figure~\ref{fig:Th_Exp},
for the call options with maturity date less or equal then ten 
working days (two weeks). We observe a coincidence inside the errors.
Finally we observe that to be able to tell a difference between
classical way of pricing derivatives  and growth-optimal
pricing procedure either the  price movement of the underlying has to have
`wilder' fluctuations, or one has to have access to more 
high-frequency data. 

\section{The general case with correlated stocks}
\label{s:multiasset}
\noindent
In order to study the effect of correlated stocks, the market in
this section is assumed to have $L$ stocks $\{S^{k}\}^{L}_{k=1}$ 
and $D_{k}$ derivatives $\{C^{(j,k)} (S^{k}_{T}) \}^{D_{k}}_{j=1}$ written on them.
To get transparency of the principle we discuss 
the case of just two stocks $L=2$ with
one derivative each and a risk free interest rate equal to $1$. 
The conclusion of this section will be given in a general notation.

The features of the stock price movements are given by:
\begin{equation}
 u_{i}^{(k)} = \frac{S^{(k)}_{i+1}}{S^{(k)}_{i}} \;\;\;\;\;\; k=1,2 \,\, .
\end{equation}
The random variable $u^{(k)}_{i}$ is assumed to be correlated on other stocks.

An investor of thgs market commits a fraction  of his capital $l_{k}$ in
stock $k$ and a fraction $d_{j,k}$ in
the derivative $j$.
The exponential growth rate of the capital is:
\begin{eqnarray}
\begin{array}{l}
 \; \lambda(l_{1},l_{2},d_{1,1},d_{1,2}) =   \\
 E^p \left[\log\left(1+l_1 (u^{(1)} -1) + l_2 (u^{(2)}-1) + 
d_{1,1} \left( \frac{C^{(1,1)}_1}{C^{(1,1)}_0} - 1 \right) +
d_{1,2} \left( \frac{C^{(1,2)}_1}{C^{(1,2)}_0} - 1 \right) \right) \right]\,\, . \\
\end{array}
\label{eq:growthgeneralcase}
\end{eqnarray}
According to the general equlibrium argument, 
one obtains the two equations:
\begin{equation}
E^q[(u^{(k)}-1)] =  0 \;\;\;\;\;\; k=1,2 
\label{eq:optimaleqstock}  
\end{equation}
and
\begin{equation}
C_{i}^{(j,k)}(S^{(k)}_i)  =  E^q[C_{i+1}^{(j,k)}(S^{(k)}_{i+1})]  
\;\;\;\;\;\; j=1,2
\label{eq:optimaleqderivative}  
\end{equation}
where we have defined, as in section~\ref{s:principle-NASA},
a new probability measure $q$:
\begin{equation}
 q^{(2)}_{h}=
\sum_m \frac{p_{hm}}{1+l^{*}_{1}(u^{(h,1)} - 1)+l^{*}_{2}(u^{(m,2)}-1)}
\label{eq:generalQ}
\end{equation}
where the index $h$ and $m$ stands for some realization of the
random variables $u^{(1)}$ and $u^{(2)}$ with the joint probability 
$p_{hm}$.

The solution of equations~(\ref{eq:optimaleqstock}) 
gives the optimal fraction $l_{k}^{*}$ to
invest in each of the different stocks $S^k$ and 
equations~(\ref{eq:optimaleqderivative}) give the option prices    
as the expected value under the measure~(\ref{eq:generalQ}).
The martingale measure may be compounded so that the option is priced
by using the pay out function at expire date:
\begin{equation}
	C_{i}^{(j,k)}(S^{(k)}_{i})=E^{Q}[C^{(j,k)}_{T}(S_{T}^{(k)})]
	\label{compmartingale}
\end{equation}
where the $Q$ denotes the compounded martingale
measure of $q$.

The case of independent stock price movements is of particular interest
from a theoretical point of view.
The price of the derivative on an underlying stock and even 
the hedge position do not depend on the other stocks in the 'classical models',
i.e. complete markets and 
risk minimization pricing procedure. 
This result is a consequence of the property
\begin{equation}
q^{(2)}_h = q_h
\label{eq:property}
\end{equation}
where $q_h$ and $q^{(2)}_h$ are the martingale measures
of one and two shares respectively.

\begin{figure}
\vspace*{13pt}
\begin{center}
\mbox{\psfig{file=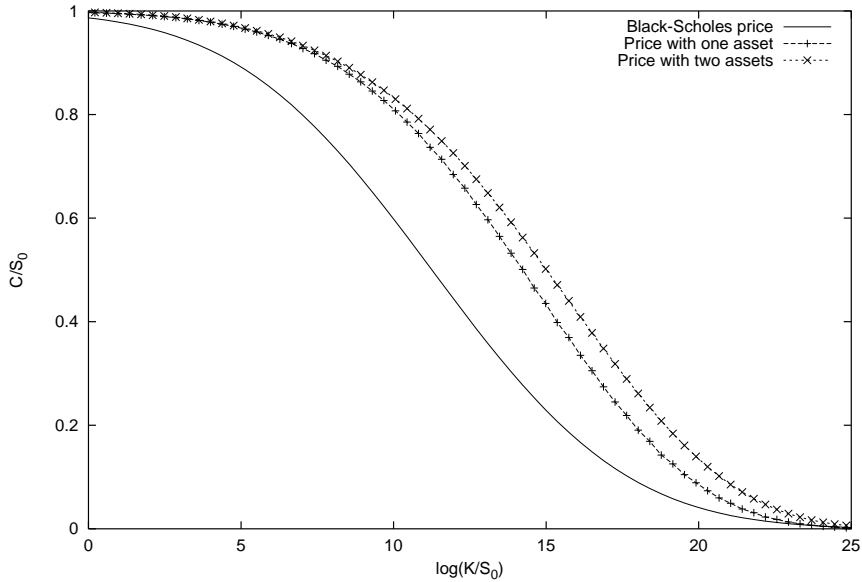,width=4.5in}}
\end{center}
\vspace*{13pt}
\caption{
Option prices $C$ rescaled by the initial value of the stock $S_0$
{\it vs.} $\log(K/S_0)$ for $T=30$ and
the returns of figure 1.  
We compare the exact prices with one and two asset in the market
with the Black-Scholes prescription.
}
\label{fig:multiasset}
\end{figure}

In the case of a complete market 
it is a known fact that the martingale
measure $q_h$ is unique~\cite{DemangeRochet}, 
then it can not
be anything else but the measure~(\ref{eq:q-definition}) 
found by optimizing the
growth rate for just that particular stock 
\begin{equation}
	q_{h}=\frac{p_{h}}{1+l^{*}(u^{(h,1)} - 1)}\,\, .
\label{eq:martingale_one_share}
\end{equation}
The property~(\ref{eq:property}) is also preserved
in the risk minimization pricing procedure but not
in the general case of an incomplete market.  
In this situation the associated 
martingale measure~(\ref{eq:generalQ})
depend on the number of stocks traded in the market
\begin{equation}
 q^{(2)}_{h}= p_{h}\cdot \sum_{m}\frac{p_{m}}
{1+l^{*}_{1}(u^{(h,1)} - 1)+l^{*}_{2}(u^{(m,2)}-1)}
\neq q_h \,\, .
\label{sumqprob}
\end{equation}
In figure~\ref{fig:multiasset} we show this difference on the option
prices for the trichotomic case. 

This fact is clear in
an incomplete market with
large number $L$ of stocks
and the returns of these stocks are independent identically distributed
random processes. 
The optimal fraction invested in each stock is $O(1/L)$
and the excess rate of return becomes almost sure.
To avoid arbitrage opportunities the
expected  excess rate of return $\mu$ must vanish 
and the measure pricing the option becomes
close to the observed probability i.e, $C \simeq E^P[C_{T}]$.
Let us notice that this corresponds to a generalization 
to a non gaussian $p(u)$
of the Bachelier theory~\cite{Bachelier}.

\section{Conclusion}
\label{s:conclusion}
We have in this paper introduced a new criterion to price
derivatives in incomplete markets derived from the theory of optimal gambling
strategies in repeated games. The criterion say that one should not
be able to construct a portfolio using derivatives that grows
almost surely at a faster exponential rate than using only the underlying
security. 

A corner-stone in modern derivative pricing theory is that
discounted prices (in friction-less markets) can be expressed as expected
pay-off, expectations taken with respect to equivalent 
martingale measures. We use the criterion  to decide which
out of many possible measures to choose.
We therefore do not need to assume a complete market to fix
unambiguously a price. 

In section~\ref{s:remarks} we compared our approach to other
procedures proposed in the literature, and in section~\ref{s:lognormal}
we discussed the lognormal limit.
In our construction we determined first the equivalent martingale
measure over one elementary time-step.
By composition we can then get the risk-neutral distribution at
finite times. In a large class of models this compounded probability
tends to the lognormal form, at the points close to the maximum,
where probability is high.
One could then conclude that we just rederive the Black-Scholes model,
except for rare events. This is however not true, because, for
us, the condition that the risk-neutral distribution is a martingale involves
also the distribution of rare and large events. These
are not lognormal.
In fact, the best lognormal approximation does not by itself have
to be a martingale, and in general differs from Black-Scholes.

\section{Acknowledgement}
\label{s:acknowledgement}
E.A. thanks Marco Avellaneda, Phil Dybvig 
and Svante Johansson for email conversations and useful remarks.
R.B. thanks Duccio Fanelli and Guglielmo Lacorata for
interesting discussions about the topic.

\end{document}